\documentclass[%
 reprint,
 amsmath,amssymb,
 aps,
]{revtex4-2}

\usepackage{graphicx}
\usepackage{dcolumn}
\usepackage{bm}

\begin{document}

\preprint{APS/123-QED}

\title{Interaction between vortex beams and diatomic molecules with rotation}

\author{Guanming Lao}
\affiliation{%
 Department of Physics and Astronomy, University of California, Los Angeles, California 90095, USA
}%

\date{\today}

\begin{abstract}
The interaction between vortex beam (VB) and molecule has drawn much attention in recent years, but the lack of theoretical method somehow limits its further analysis, especially when the molecular rotational degree of freedom is involved and coupled with the molecular electronic states. To incorporate the molecular rotation into the theoretical study, in this paper, we describe the diatomic molecular states in Hund's coupling basis and express interaction Hamiltonian in form of spherical harmonics expansion, and then investigate the rotational transition of molecular states driven by VB. The theory clearly illustrates that each photon of VB may carry a total angular momentum of 0, $\hbar$, or 2$\hbar$, and therefore could drive O, P, Q, R and S branches of diatomic molecular rotational transitions with some specific selection rules. These results indicate that VB could provide new methods for preparing and measuring the diatomic molecular states. 
\end{abstract}

\maketitle

\section{Introduction}
 The interaction of laser with matter is of fundamental interest, and nowadays the study under the quantum framework has become a flourishing field. For a vortex beam (VB) of Laguerre-Gaussian mode, which has an azimuthal phase factor $e^{il\phi}$, it is shown that each photon can averagely carry orbital angular momentum (OAM) of $l\hbar$ as well as spin angular momentum (SAM) of $\pm\hbar$ \cite{allen_orbital_1992}, where $l$ is an integer and refers to the topological charge of the field. To fully understand the mechanical and quantum effects of the angular momentum carried by the vortex beams, over the past few decades, a lot of studies have been proposed about the interactions between vortex beams and various kinds of materials or objects, such as exchanging OAM in quantum-dots \cite{mahdavi2020manipulation}, particle trapping and manipulation \cite{gahagan_optical_1996, shvedov2010optical, radozycki_manipulating_2019, liu2015trapping, liu2013optical}, and driving atomic transitions \cite{ schmiegelow_light_2012, mukherjee_interaction_2018, andersen_quantized_2006, babiker_atoms_2019}. 
 
It is well-known that the SAM of light could be transferred to atoms in the atom-light dipole interaction, and the change of the atomic angular momentum and magnetic quantum number can be described by the selection rules. However, when a vortex beam of Laguerre-Gaussian (LG) mode interacts with an atom, the selection rules involving OAM are different. By expressing the dipole and quadrupole effects of the interaction as the zero- and first-order Hamiltonian, respectively, the theory shows that the transfer of OAM occurs between the internal motion and the light beam in form of quadratic effects, while the OAM does not engage with the electronic motion in dipole active transition \cite{babiker_atoms_2019, lembessis_enhanced_2013, babiker_orbital_2002}. These predictions have been corroborated in several experiments, such as the observation of the $4^2S_{1/2}\leftrightarrow3^2D_{5/2}$ quadrupole transition of trapped $^{40}\textrm{Ca}^+$ ions driven by laser beams of different LG modes \cite{schmiegelow2016transfer}, and an attempt of verifying whether the OAM of light could have the same ability and manner of driving the atomic transition as occurs for the SAM \cite{giammanco2017influence}. 

The match of the theoretical and experimental studies about the interaction of vortex beams and the bound electron in atoms inspires people to exploit the potential of the OAM. Currently, some attempts have been made towards its interaction with the bounded electron in some other complex systems. For example, the interaction between LG vortex beams and H$_2^+$ and HD$^+$ molecules has been study theoretically \cite{mondal_angular_2014}. Since the diatomic molecular system contains two nuclei which break the spherical symmetry of the electronic wave function, the electronic orbital angular momentum $\textit{\textbf{L}}$ is no longer conserved; in addition, the broken spherical symmetry and the size of the molecule in different dimensions would also introduce the rotational structure in the molecular energy levels. 

Therefore, in order to discuss the physics of vortex beam-molecule (VB-molecule) interaction with more details, the study should be proposed in a set of suitable eigenbasis of the molecular wave functions. Although Ref.\cite{mondal_angular_2014} has explained the VB-molecule interaction with a clear picture, the electronic wave function basis used for the H$_2^+$ model is not good enough for describing the molecule with complex electronic structure. Since the transfer of OAM could change the total angular momentum of the molecule, the rotational angular momentum of the nuclei framework could also be changed during the VB-molecule interaction. In this paper, to figure out how the angular momentum of the molecule is changed during the interaction, we choose the basis of Hund's case(a) and (b) to describe diatomic molecular states.

This paper is organized as follows. In Section \ref{Mod}, the model for VB-molecule interaction is developed. Then we expand the interaction Hamiltonian in spherical harmonics in Section \ref{TheosubA}, calculate the transition amplitudes between different molecular states and discuss the selection rules in Section \ref{TheosubB}. For further explanation, in Section \ref{TheosubC}, we simulate the rotational spectrum of strontium monofluoride (SrF) in $B^2\Sigma^+\leftarrow X^2\Sigma^+$ system at about 580nm. Section \ref{DiscA} briefly discusses the cases of when VB interacts with a larger molecule or the functional groups on a molecule, and Section \ref{DiscB} discusses the interaction between molecule and non-canonical vortex beams, which contain phase factors different from $e^{il\phi}$. Section \ref{Conc} concludes the paper.    
\section{Model}
\label{Mod}

\begin{figure}
    \centering
    \includegraphics[scale=0.35]{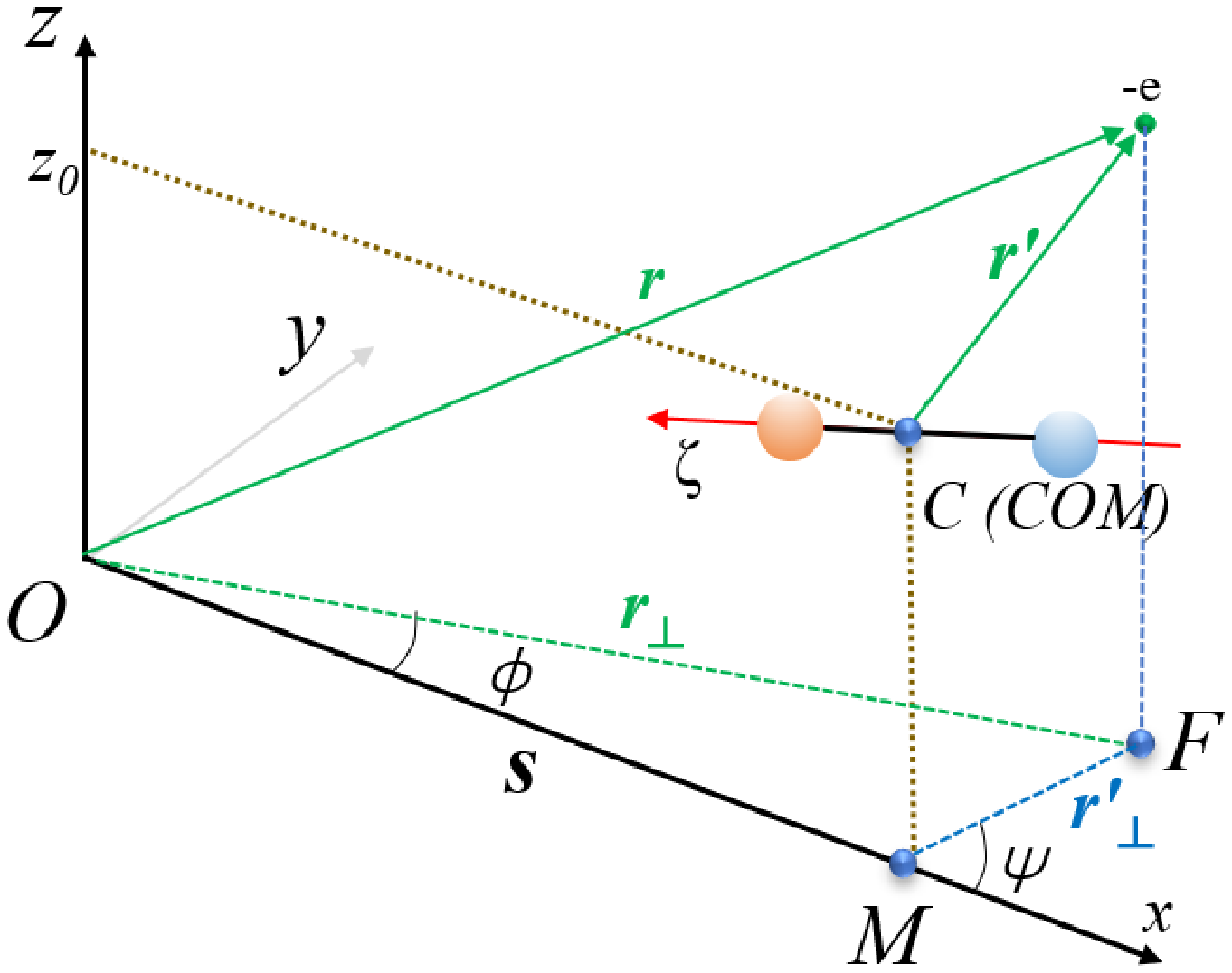}
    \caption{Geometric definitions. The beam is propagating in $z$ direction, plane $xOy$ is the transverse plane, with the phase factor $e^{il\phi}$ depending on the azimuthal angle $\phi=\angle FOx$. The COM of the molecule, denoted as point $C$, is located at $(s, 0, z_0)$, and the position vector of the bounded electron $-e$ is $\textit{\textbf{r}}=(r_\perp, \phi, z)$ in the cylindrical coordinates. The origins of space- and molecule-fixed spherical coordinates are both at $C$, with polar axis to be $z$ and the molecular axis $\zeta$, respectively. Point F and M are respectively the projection of the electron position and C on the $xOy$ plane. $\textit{\textbf{r}}'$ is the position vector of electron with respect to C, and its angular coordinates $\psi$ and $\theta$ in the space-fixed spherical coordinates are respectively defined as $\angle FMx$ and the polar angle about $z$ axis. In the molecule-fixed spherical coordinates, $\theta'$ is the polar angle about $\zeta$ axis, and $\psi'$ could be well defined if a zero position is given.}
    \label{OMOA}
\end{figure}

To start, we first consider the Hamiltonian for the interaction between a diatomic molecule and a circular polarized VB propagating along $+z$ direction. The optical field in space-fixed cylindrical coordinates $(r_\perp, \phi, z)$ reads
\begin{equation}
   \mathbf{E^{\pm}}=E(r_\perp, \phi, z)\hat{\epsilon}^{\pm}= F(r_\perp, z)e^{il\phi}\hat{\epsilon}^{\pm},
\end{equation}
where $\hat{\epsilon}^+$ and $\hat{\epsilon}^-$ (as well as the $+$ and $-$ signs in the superscript of $\mathbf{E}$) denote the left- and right-handed circular polarization of the beam, respectively, $F(r_\perp, z)$ is the field amplitude distribution in space, $e^{il\phi}$ is the helical phase factor. The temporal factor $e^{-i\omega_0 t}$ is omitted for simplicity. The total interaction Hamiltonian is mainly contributed by dipole and quadrupole interaction terms, i.e.,
\begin{equation}
    H_{int}=H_d+H_{quad}.
\end{equation}
The dipole interaction Hamiltonian reads
\begin{equation}
\label{Hd}
    H_d=\mathbf{E}^{\pm}\cdot \mathbf{d}=E(r_\perp, \phi, z)\hat{\epsilon}^{\pm}\cdot\mathbf{d},
\end{equation}
here $\mathbf{d}$ is the molecular dipole moment. Since the optical vortices are usually carried by lasers in visible, IR, or UV ranges, which are having enough energy to drive molecular electronic transition, we will take the molecular electronic transition into our consideration. For simplicity, we assume that the field is mainly interacting with one bound electron of the molecule by driving it to a new electronic state, and the distance between the nuclei is not changed significantly during the interaction so that the the dipole moment of the nuclei as well as the electrons not interacting with the field could be ignored. Therefore, by defining $\textit{\textbf{r}}'$ as the bound electron's position vector relative to the center of mass (COM) of the molecule (see Fig.\ref{OMOA}), we may use the electron's dipole moment operator $-e\textit{\textbf{r}}'$ to replace $\mathbf{d}$, as an approximation. 

In a given Cartesian coordinate system $(r_1, r_2, r_3)$, the expression of quadrupole interaction Hamiltonian is
\begin{equation}
    H_{quad}=\frac{1}{2}\nabla_i E_j^{\pm} Q_{ij}
\end{equation}
where $i, j=1, 2, 3$, $\nabla_i=\partial/\partial r_i$ is the $i$ component of the gradient, $E_j^{\pm}$ is the $j$ component of electric field,  $Q_{ij}$ denotes the $ij$ component of the molecular quadrupole moment, and the Einstein summation convention is applied. 

For the diatomic molecular states, here we express it in the Hund's case(a) basis \cite{lefebvre-brion_spectra_2004, brown_rotational_nodate}
\begin{equation}
    |\eta, \Lambda; J, \Omega, M_J, S, \Sigma\rangle=|\eta \Lambda\rangle|J \Omega M_J\rangle|S \Sigma\rangle,
\end{equation}
where $\eta$ denotes all the other quantum numbers not listed (vibration, electronic configuration, etc.), $\Lambda$ is the projection of the electronic angular momentum $L$ on the molecular axis $\zeta$, $J$ is the total angular momentum of the molecule, $M_J$ and $\Omega$ are the projection of $J$ on the space-fixed $z$ axis and the internuclear axis $\zeta$, respectively, and $S$ and $\Sigma$ are respectively the total spin angular momentum of the electrons and its axial component. The wave function on the right-hand side have their own meaning: $|\eta \Lambda\rangle$ is the electronic wave function, $|J \Omega M\rangle$ is the rotational wave function for symmetric top (see Appendix \ref{helib}), and $|S \Sigma\rangle$ is the wave function for electronic spin. The nuclear spin $I$, which mainly contributes to the molecular hyperfine structure, is not considered in this paper for simplicity. 
Note that rotational wave function $|J \Omega M_J\rangle$ is also the common eigenvector of operators $\textbf{\textit{J}}^2$, $J_{\zeta}$ and $J_{z}$ (set $\hbar=1$): 
 \begin{eqnarray}
     &&J^2|J \Omega M_J\rangle=J(J+1)|J \Omega M_J\rangle,\\
     &&J_{\zeta}|J \Omega M_J\rangle=\Omega|J \Omega M_J\rangle,\\
     &&J_{z}|J \Omega M_J\rangle=M_J|J \Omega M_J\rangle.
 \end{eqnarray}
When the interaction Hamiltonian $H_{int}$ drives transition $|\eta', \Lambda', J', \Omega', M_{J'}\rangle\leftarrow|\eta, \Lambda, J, \Omega, M_J\rangle$,  the corresponding transition amplitude reads:
\begin{equation}
\label{eq4}
    \langle \eta', \Lambda', J', \Omega', M_{J'}|H_{int}|\eta, \Lambda, J, \Omega, M_J\rangle.
\end{equation}
where the primed and unprimed quantum numbers are for labeling the initial and final states, respectively, and spin wave function is neglected because $H_{int}$ does not directly affect the electronic spin state. 

However, to calculate the transition amplitude, there is still one problem standing on the way. Currently the electric field of VB is expressed in space-fixed cylindrical coordinates $(r_\perp, \phi, z)$, while the molecular states $|\eta \Lambda\rangle$ and $|\eta' \Lambda'\rangle$ are conventionally described in the molecule-fixed axis system (here we use spherical coordinates $(r', \theta', \psi')$). To keep going, it is necessary to express the Hamiltonian in the same coordinate representation as the molecular states. To achieve this, we will first express the field in the space-fixed spherical coordinates $(r', \theta, \psi)$ with COM as its origin and $z$ axis as its polar axis, and then convert it to the molecule-fixed spherical coordinates $(r', \theta', \psi')$ with rotation operator $D(\omega)$ \cite{lefebvre-brion_spectra_2004, brown_rotational_nodate}. After the conversion, we will calculate the transition amplitude and discuss the rotational transition driven by the VB. 

\section{Theory}
\label{Theo}
\subsection{Spherical harmonics expansion of interaction Hamiltonian}
\label{TheosubA}
It is necessary to confine the molecule in the vicinity of the optical vortex for obtaining convincing experimental signals \cite{schmiegelow2016transfer}. However, there could still be some distance $s$ towards the central axis of the vortex beam, even though the combination of laser cooling  \cite{anderegg_laser_2018, zhelyazkova_laser_2014, hudson2020laser, kozyryev_sisyphus_2017, lev2008prospects, shuman_laser_2010} and ion trap or Magneto-optical trap (MOT) could be applied for suppressing the undesired motion and confine the molecules \cite{barry2014magneto, schwarz2012cryogenic, mccarron2015improved, stollenwerk2019ip}. For instance, according to the averaged velocity of the cooled SrF molecule $v\sim 1 \mathrm{m s^{-1}}$ and the spontaneous decay rate $\Gamma=2\pi\times7\mathrm{MHz}$ \cite{shuman_laser_2010}, one can estimate that the SrF molecule could be confined in a scale of 10-100nm in the MOT. Since such dimensional scale is much larger than the size of the diatomic molecule (usually of a few Angstroms), we can expect that the diatomic molecule could be treated as a small rotor in the helical optical field $E(r_\perp, \phi, z)$, and therefore it is reasonable to base the theory of this paper on the assumption of $s\gg r'$. 

We would start with the following geometric relations in Fig.\ref{OMOA}:
\begin{eqnarray}
     &&r_\perp=\sqrt{s^2+r_\perp^{'2}+2sr'_\perp \mathrm{cos}\psi},\\
     &&\phi=\mathrm{arctan}\frac{\mathrm{sin}\psi}{\mathrm{cos}\psi+s/r'_\perp},
\end{eqnarray}
where $r'_\perp$ is the projection of position vector $\textit{\textbf{r}}'$ on the $xOy$ plane. Since the bound electron is very close to the COM, the approximation $r'_\perp=r'\mathrm{sin}\theta\ll s$ is valid, and thus 
\begin{equation}
\label{aprx}
    \phi \approx \frac{\mathrm{sin}\psi}{\mathrm{cos}\psi+s/r'_\perp}\approx\frac{r'_\perp\mathrm{sin}\psi}{s}, r_\perp \approx s+r'_\perp\mathrm{cos}\psi.
\end{equation}
Therefore, the helical electric field in the vicinity of the molecule could be expressed in the space-fixed spherical coordinate as follows:
\begin{equation}
\label{Eapprox}
\begin{split}
    E&(r_\perp, \phi, z)\approx \bigg[F(s, z_0)+\left.\frac{\partial F}{\partial r_\perp}\right|_{r_\perp=s}\Delta r_\perp\\
    &+\left.\frac{\partial F}{\partial z}\right|_{z=z_0}(z-z_0)\bigg]\mathrm{exp}\left(il\frac{r'_\perp\mathrm{sin}\psi}{s}\right)\\
    \approx&[F_0+F_\alpha r'_\perp\mathrm{cos}\psi+F_\beta r'\mathrm{cos}\theta]\left(1+il\frac{r'_\perp\mathrm{sin}\psi}{s}\right)\\
    \approx&F_0+r'\mathrm{sin}\theta\left[F_\alpha \mathrm{cos}\psi+iF_0l\frac{\mathrm{sin}\psi}{s}\right]+F_\beta r'\mathrm{cos}\theta\\
    =&F_0+r'\displaystyle\sum_{p=-1}^1 C_p Y_1^p(\theta,\psi),
\end{split}
\end{equation}
where
\begin{eqnarray}
&&F_0=F(s, z_0),\\
&&F_\alpha=\left.\frac{\partial F(r_\perp, z_0)}{\partial r_\perp}\right|_{r_\perp=s},\\  &&F_\beta=\left.\frac{\partial F(s, z)}{\partial z}\right|_{z=z_0},\\
&&C_1=-\sqrt{\frac{2\pi}{3}}\left[F_\alpha+F_0\frac{l}{s}\right],\\
&&C_0=\sqrt{\frac{4\pi}{3}}F_\beta,\\
&&C_{-1}=\sqrt{\frac{2\pi}{3}}\left[F_\alpha-F_0\frac{l}{s}\right].
\end{eqnarray}
Here we only keep the zeroth and first order terms of $r'$ in the electric field expression  Eq.(\ref{Eapprox}). The zeroth order term $F_0$ is the intensity of the electric field at the molecular COM, and the first order term of the field is expressed as a linear combination of $Y_1^p(.)$, which describes the spatial variation of the field. Note that the OAM is originated from the field spatial distribution, the dipole interaction involving the first order of $r'$ in the field amplitude generates one quanta of OAM (because the ranks of these spherical functions in the expression are 1). This is in agree with the theoretical study Ref.\cite{babiker_orbital_2002}. 

In the interaction Hamiltonian, the transfer of the photonic SAM is decided by the polarization of the field, i.e., $\mathbf{\hat{\epsilon}^{\pm}\cdot d}$. In the helicity basis (see Appendix \ref{helib}) of the lab frame, this term could be expressed as follows:
\begin{equation}
\label{poldotd}
    \mathbf{\hat{\epsilon}^{\pm}\cdot d}=-er'\sqrt{\frac{4\pi}{3}}Y_1^{\pm1}(\theta, \psi).
\end{equation}
Obviously, the electric field in the dipole interaction Hamiltonian Eq.($\ref{Hd}$) is contributed by the constant part $F_0$ as well as the spherical parts in Eq.(\ref{Eapprox}). To distinguish their contribution to the dipole interaction, we define the interaction induced by the constant part $F_0$ as the first order dipole interaction, and the product of the dipole moment and the spherical part as the second order dipole interaction. The first order dipole interaction induces the well-know dipole transition of the molecular states, with one quanta of SAM transferred between photon and molecule. In the second order dipole interaction, one quanta of OAM is transferred between the molecule and vortex beam accompanied with the SAM, which is in agree with the conclusion of Ref.\cite{babiker_orbital_2002}. Though we could expect that higher orders terms omitted in the field expression Eq.(\ref{Eapprox}) have spherical harmonics of higher ranks and therefore could bring more OAM into the interaction, for simplicity, we only focus on the effects of the first and second order terms in the VB-molecule interaction Hamiltonian, and we assume that the effects of high order terms are negligible. 

On substitute Eq.(\ref{Eapprox}) and Eq.(\ref{poldotd}) into Eq.(\ref{Hd}), the dipole interaction Hamiltonian could be expressed as below:
\begin{equation}
\label{EY1}
\begin{split}
    H_d\approx& -\sqrt{\frac{4\pi}{3}}er'\bigg[F_0Y_1^{\pm1}(\theta, \psi)\\
    &+r'\displaystyle\sum_{k=0}^2\sum_{p=-k}^k\sum_{p_1=-1}^1\sqrt{\frac{9(2k+1)}{4\pi}}(-1)^pC_{p_1}\\
    &\times\left(\begin{matrix}1&1&k\\ p_1&\pm1&-p\end{matrix} \right)\left(\begin{matrix}1&1&k\\0&0&0\end{matrix} \right) Y_{k}^{p}(\theta, \psi)\bigg],
\end{split}
\end{equation}
where $(:::)$ is the Wigner 3-j symbol and the formula for the product of spherical harmonics functions \cite{brown_rotational_nodate} has been used. The first term of Eq.(\ref{EY1}) on the right hand side is the first order dipole interaction in which the photon has just one quanta of SAM, and the second term is the second order dipole interaction. For non-zero results of the second 3-j symbol in the last line, $k$ can only be 0 or 2, which indicates that, in the second order dipole interaction, the sum of one quanta of SAM and one quanta of OAM forms the photon's total angular momentum of 0 or $2\hbar$. Therefore, in the second order interaction, the total angular momentum transferred between the photon and molecule could only be 0 or $2\hbar$, or such transition amplitude is 0. This is also in agree with the conclusion of Ref.\cite{babiker_orbital_2002}.

The two possible values of the photon's total angular momentum could lead to two branches of transitions. The branch of $k=2$ is corresponding to a case in which a VB photon can carry two quanta of angular momentum, which is verified by the experiment results in Ref.\cite{schmiegelow2016transfer}. The branch of $k=0$ is corresponding to a case in which the VB photon could drive the dipole forbidden transitions that maintain the quantum number of $J$ and the parity of the molecular states. However, currently, no relative experiment scheme has been proposed yet. 

For quadrupole interaction $H_{quad}$, the Hamiltonian reads
\begin{equation}
\label{Hquad}
\begin{split}
    H_{quad}&=\frac{1}{2}\nabla_i E_j Q_{ij}=\frac{1}{2}T^2(\nabla \mathbf{E})\cdot T^2(Q)\\
    &=\displaystyle\sum_{p=-2}^2\frac{(-1)^p}{2}T_p^2(\nabla \mathbf{E})T_{-p}^2(Q),
\end{split}
\end{equation}
here we use $T^2(.)$ notation to represent the rank-2 tensor expression of the operators. For quadrupole moment Q, its expression in space-fixed coordinates is 
\begin{equation}
    T_p^2(Q)=(-e)\sqrt{\frac{16\pi}{5}}r'^2Y_2^p(\theta, \psi).
\end{equation}
For field gradient operator $\nabla E$, since we only consider the quadrupole interaction which is proportional to $r'^2$ and would omit the higher order terms, and the quadrupole moment $Q$ has already contained the term of $r'^2$, $T_p^2(\nabla \mathbf{E})$ could be approximated to some constants. After some calculation, the total Hamiltonian $H_{int}$ could be expressed as a sum of spherical harmonics:
\begin{equation}
    H_{int}\approx\displaystyle\sum_{k=0}^2\sum_{p=-k}^k a_{k,p}^{(\pm)}r'^{n_k}Y_k^p(\theta, \psi),
\end{equation}
where $n_k=1$ when $k=1$ and $n_k=2$ for the other two $k$ values, and the expressions of the coefficients $a_{k,p}^{(\pm)}$ could be found in Appendix \ref{ApB}.

\subsection{Calculation of transition matrix elements}
\label{TheosubB}
After having expressed the interaction Hamiltonian $H_{int}$ as the sum of spherical harmonics, the calculation of the transition matrix elements Eq.(\ref{eq4}) could be attributed to the calculation of the matrix elements of $r^{n_k}Y_k^p(\theta, \psi)$ in the basis of molecular states: 
\begin{equation}
\label{TrY}
\begin{split}
    &\langle \eta', \Lambda'; J', \Omega', M_{J'}|H_{int}|\eta, \Lambda; J, \Omega, M_J\rangle\\
    =&\sum_{k, p} a_{k,p}^{(\pm)}\langle \eta'\Lambda'|\langle J'\Omega'M_{J'}|r^{n_k}Y_k^p(\theta, \psi)|J\Omega M_J\rangle|\eta\Lambda\rangle.
\end{split}
\end{equation}
By applying rotational operation on the spherical function $Y_k^p(\theta, \psi)$ and expressing it in molecule-fixed angular coordinates $(\theta', \psi')$, one can obtain
 \begin{equation}
     Y_k^p(\theta, \psi)=\sum_{q=-k}^kD_{pq}^{k}(\omega)^*Y_k^{q}(\theta', \psi'),
 \end{equation}
where $D_{pq}^{k}(\omega)^*$ is the complex conjugate of the $pq$ element of the $k$th rank rotational matrix $\mathcal{D}^{(k)}(\omega)$, with $p$ the space-fixed tensor components and $q$ the molecule-fixed components, and $\omega$ is the Euler angles through which the $z$ axis coincide with the $\zeta$ axis under the rotation (same definition as Appendix \ref{helib}). Noticed that $D_{pq}^{k}(\omega)^*$ is the coefficient of tensor components and the rotational wave functions are also the functions of $\omega$, while $r'^{n_k}Y_k^{q}(\theta', \psi')$ is the tensor component of $H_{int}$ and therefore should act on the electronic state $|\eta\Lambda\rangle$, one can calculate each term in Eq.(\ref{TrY}) as follows \cite{brown_rotational_nodate}:
\begin{equation}
\begin{split}
\label{rnYpk}
    &\langle \eta'\Lambda'|\langle J'\Omega'M_{J'}|r^{n_k}Y_k^p(\theta, \psi)|J\Omega M_J\rangle|\eta\Lambda\rangle\\
    =&\displaystyle\sum_{q=-k}^k(-1)^{J'-M_{J'}}\left(\begin{matrix}J'&k&J\\-M_{J'}&p&M_J\end{matrix} \right)(-1)^{J'-\Omega'}\\
    &\times[(2J'+1)(2J+1)]^\frac{1}{2}\left(\begin{matrix}J'&k&J\\-\Omega'&q&\Omega\end{matrix} \right)\\
    &\times\langle\eta'\Lambda'|r'^{n_k} Y_k^{q}(\theta', \psi’)|\eta\Lambda\rangle,
\end{split}
\end{equation}
For non-zero results of the two 3-j symbols, the following selection rules under the space-fixed and molecule-fixed frames should be satisfied:
\begin{eqnarray}
    &&\Delta M_J=M_{J'}-M_J=p,\\
    &&\Delta \Omega = \Omega'-\Omega=q,\\
    \label{Tranrule}
    &&J'=J+k, J+k-1, \dots, |J-k|, k=0,1,2.
\end{eqnarray}
Here the first two equations are the selection rules for the magnetic quantum numbers $M_J$ and $\Omega$, and the last one is the selection rule for the angular momentum $J$. Note that, in Eq.(\ref{rnYpk}), the terms $\langle\eta'\Lambda'|r'^{n_k} Y_k^{q}(\theta', \psi’)|\eta\Lambda\rangle$ are just some constants if the initial and final electronic states are given, and therefore do not affect the selection rules. 

Now the selection rules as well as the transition amplitude formula Eq.(\ref{TrY}) could help us to find available transitions. For example, consider the quadrupole transition $\Delta J=J'-J=\pm2$ of the diatomic molecule described in Hund's case(a) basis $|\eta,\Lambda;J, \Omega, M_J, S, \Sigma \rangle$. These two branches could only be driven by the $k=2$ terms in Eq.(\ref{TrY}), so one can calculate the transition amplitude Eq.(\ref{TrY}) as follows
\begin{equation}
\label{qtca}
    \begin{split}
    &\langle \eta',\Lambda'; J', \Omega', M_{J'}, S', \Sigma'|H_{int}|\eta,\Lambda; J, \Omega, M_J, S, \Sigma\rangle\\
    =&\displaystyle\sum_{p=-2}^2\sum_{q=-2}^2\delta_{S'S}\delta_{\Sigma' \Sigma}a_{2,p}^{(\pm)}\langle\eta'\Lambda'|r'^2Y_2^{q}(\theta', \psi’)|\eta\Lambda\rangle(-1)^{M-\Omega}\\
    &\times[(2J'+1)(2J+1)]^\frac{1}{2}\left(\begin{matrix}J'&2&J\\-M_{J'}&p&M_J\end{matrix} \right)\left(\begin{matrix}J'&2&J\\-\Omega'&q&\Omega\end{matrix} \right)\\
    =&\sum_{p=-2}^2\delta_{S'S}\delta_{\Sigma' \Sigma}a_{2,p}^{(\pm)}\langle\eta'\Lambda'|r'^2Y_2^{0}(\theta', \psi’)|\eta\Lambda\rangle(-1)^{M-\Omega}\\
    &\times[(2J'+1)(2J+1)]^\frac{1}{2}\left(\begin{matrix}J'&2&J\\-M_{J'}&p&M_J\end{matrix} \right)\left(\begin{matrix}J'&2&J\\-\Omega'&0&\Omega\end{matrix} \right).
\end{split}
\end{equation}
The last step which eliminates the summation over $q$ is explained in Appendix \ref{helib}. By knowing which transitions are available, we can apply it to some further study, such as spectroscopy.
\subsection{Simulation of SrF rotational spectroscopy}
\label{TheosubC}
After the calculation of the transition amplitude, one can learn the allowed transitions of a certain molecule and then study its rotational spectroscopy. For illustration, we will calculate the transition amplitudes for the ($v'=0\leftarrow v=0$) rotational band of $B^2\Sigma^+ \leftarrow X^2\Sigma^+$ system of SrF molecule \cite{steimle1977rotational} and simulate the excitation spectra of the system. Since the spin-orbital coupling effect is absent in both the ground molecular electronic state $X^2\Sigma^+$ and the excited state $B^2\Sigma^+$, and the spin-rotational coupling effect mixes different spin states $|S, \Sigma\rangle$ in the Hund's case(a) basis, Hund's case(b) basis $|\eta, \Lambda; J, N, S, M_J\rangle$ is a better choice for the study. The relation between the case(a) and (b) basis is:
\begin{figure}
    \centering
    \includegraphics[scale=0.6]{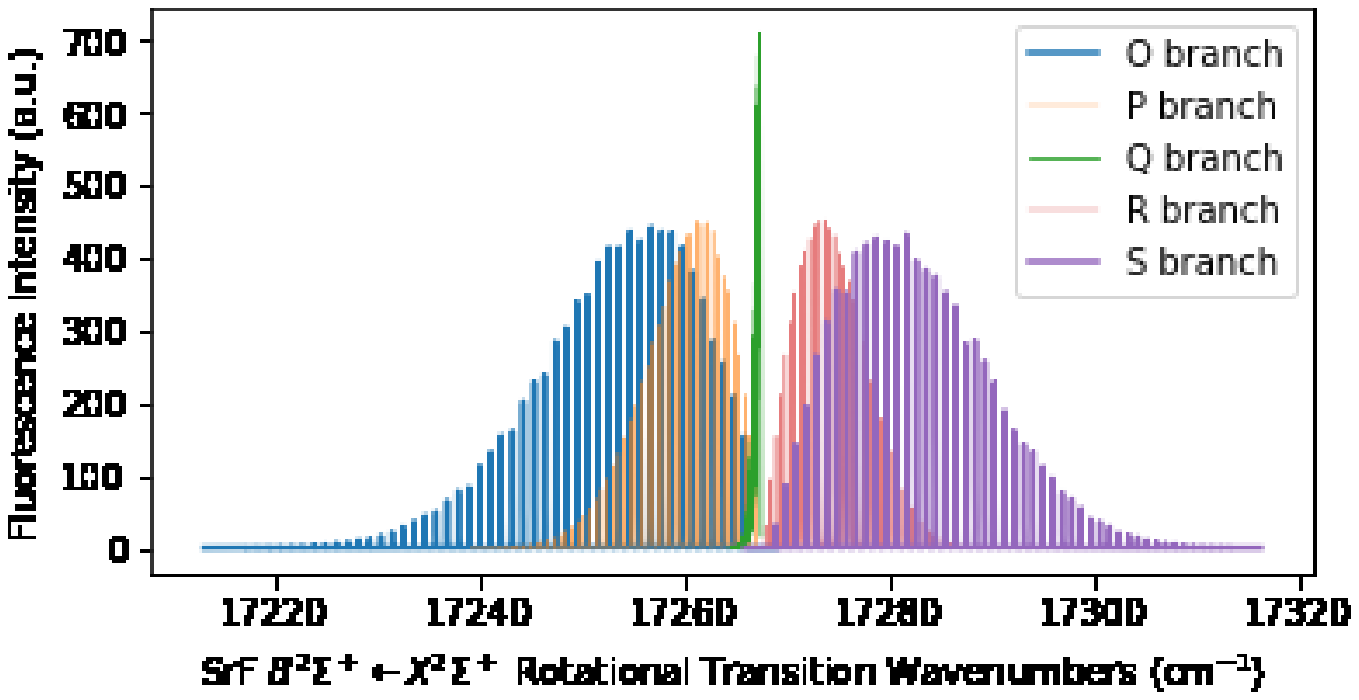}
    \caption{Simulation example of SrF excitation spectrum at temperature of rotation $T_r=100K$. P and R branches are the lines in orange and red, respectively, corresponding to the dipole transition, and O, Q, R branches (lines in blue, green, and purple) are corresponding to the lines of quadrupole transition and the transition with OAM. The lines in Q branch ($\Delta N=0$) pile up because the difference between the rotational constants $B_X$ and $B_B$ is small. The line intensity is proportional to the  ground states population, which is proportional to the degeneracy of rotational states, i.e., $2J+1$, as well as $\textrm{exp}(-H_{Rot}/k_BT_r)$, the factor given by the Boltzmann distribution.}
    \label{SrF}
\end{figure}
\begin{equation}
\label{cB2A}
\begin{split}
    |\eta, \Lambda; J, N, &S, M_J\rangle=\displaystyle\sum_{\Sigma=-S}^S(-1)^{J-S+\Lambda}\sqrt{2N+1}\\
    &\times\left(\begin{matrix}J&S&N\\ \Omega&-\Sigma&-\Lambda\end{matrix} \right)|\eta,\Lambda;J, \Omega, M_J, S, \Sigma \rangle,
\end{split}
\end{equation}
where $\textbf{\textit{N}}=\textbf{\textit{J}}-\textbf{\textit{S}}$ is the total angular momentum without spin. The transition amplitude in Hund's case(b) basis reads:
\begin{equation}
\label{cbtr}
\begin{split}
    &\langle\eta', \Lambda'; J', N', S', M_{J'}|H_{int}|\eta, \Lambda; J, N, S, M_J\rangle\\
    =&\displaystyle\sum_{\Sigma=-S}^S(-1)^{J-J'+\Lambda-\Lambda'}\sqrt{(2N+1)(2N'+1)}\\
    &\times\delta_{S'S}\delta_{\Sigma' \Sigma}\left(\begin{matrix}J'&S&N'\\ \Omega&-\Sigma&-\Lambda'\end{matrix} \right)\left(\begin{matrix}J&S&N\\ \Omega&-\Sigma&-\Lambda\end{matrix} \right)\\
    &\times\langle \eta',\Lambda'; J', \Omega', M_{J'}, S, \Sigma|H_{int}|\eta,\Lambda; J, \Omega, M_J, S, \Sigma\rangle,
\end{split}
\end{equation}
where the conditions for non-zero values of Eq.(\ref{qtca}) were used, and the rest of the calculation could be attributed to the case(a). This equation actually indicates that the amplitudes of transitions with $\Delta N=N'-N=0, \pm2$ may have non-zero values, and therefore the transitions may be allowed. For example, consider the transition between two different quantum number sets $(J=1/2, N=0, \Lambda=0)$ and $(J'=5/2, N'=2, \Lambda'=0)$, respectively. By calculating with all these values, the transition amplitude is proportional to
\begin{equation}
\begin{split}
    &\displaystyle\sum_{\Sigma=-S}^S(-1)^{J-J'+\Lambda-\Lambda'}\left(\begin{matrix}J'&2&J\\-\Omega&0&\Omega\end{matrix} \right)\\
    &\times\left(\begin{matrix}J'&S&N'\\ \Omega&-\Sigma&-\Lambda'\end{matrix}\right)\left(\begin{matrix}J&S&N\\ \Omega&-\Sigma&-\Lambda\end{matrix} \right) =\frac{\sqrt{10}}{10}\neq 0,
\end{split}
\end{equation}
which shows that the transition between the states $|B^{2}\Sigma, \Lambda'=0; J'=5/2, N'=2, S, M_J\rangle \leftarrow |X^{2}\Sigma, \Lambda=0; J=1/2, N=0, S, M_J\rangle$ is allowed in the VB-molecule interaction. 

After calculating Eq.(\ref{cbtr}) for different pairs of initial and final states, one can sort out the allowed rotational transitions and carry out the study of rotational spectroscopy. The molecular rotational Hamiltonian reads
\begin{equation}
    H_{Rot}=B(v)N(N+1),
\end{equation}
here $v$ is the vibrational quantum number and $B(v)$ is the rotational energy constants. For $X$ and $B$ vibrational ground states, the rotational constants are $B_X=0.25075\text{cm}^{-1}$ and $B_B=0.24961\text{cm}^{-1}$, respectively \cite{steimle1977rotational}. The transition energies could therefore be calculated as follows
\begin{equation}
\begin{split}
    \Delta E=&E_B(v, N') - E_X(v, N)\\
    =&T_e+B_BN'(N'+1)-B_XN(N+1),
\end{split}
\end{equation}
where $T_e=17267.42\text{cm}^{-1}$ is the electronic transition energy. With the given constants, the excitation spectra could be studied by simulation, as shown in Fig.\ref{SrF}. This is to simulate the experiment of illuminating the SrF molecule with high intensity laser pulses with vortex phase distribution $e^{i\phi}$ at 580nm. Since the high intensity vortex beam pulses can generate great field gradient in the vicinity of the center, it can increase both the dipole and quadrupole transition rates significantly and transfer almost all the ground state molecules to the available excited states, which then emit photons for state detection. Hence the excitation spectra comprise five branches of transitions: O branch with $\Delta N=-2$, P branch with $\Delta N=-1$, Q branch with $\Delta N=0$, R branch with $\Delta N=1$, and S branch with $\Delta N=2$. Accompanied with the analysis in the first subsection, we can know that as the transition is driven by VB, the first order dipole interaction of the optical field can drive the transitions of P and R branches, the $k=0$ branch of second order dipole interaction allows the Q branch transition, and the quadrupole transition as well as the $k=2$ branch of the second order dipole interaction allow the O and S branches. This shows the OAM carried by the VB photon could enable extra rotational transitions and therefore provide us more methods for preparing molecular rotational states.
\section{Discussion}
\label{Disc}
\subsection{Interaction with larger molecules}
\label{DiscA}
Applying VB to drive extra rotational lines of diatomic molecule naturally leads to a generalized idea: VB may also be able to drive rotational transitions of some larger molecules with optical cycling properties, such as molecules of alkaline-earth monoalkoxides type \cite{brazier1986laser} (which could be studied with the symmetry top rotational wave function), or big organic molecules with several functional groups having bound electrons active for the transition. Though it may be hard to theoretically analyze the rotational spectroscopy of the latter type of molecules since they are usually treated as asymmetry rotors and thus having complicated rotational eigen-wavefunctions, we may still be able to make use of the VB-molecule interaction and bring out theoretical analysis in some specific situations.

For concreteness, consider a molecule having two functional groups, $G_1$ and $G_2$, with each group has an optical cycling center available for VB-molecule interaction. The groups and the rest of the molecule are connected by single bonds, and therefore these groups could have rotational degree of freedom about the axes lying on the bonds. We can assume that each group has angular momentum 
\begin{equation}
    J_i=R_i+L_i+S_i, ~i=1, 2,
\end{equation}
where $R_i, L_i$ and $S_i$ are the rotational, electronic orbital, and electronic spin angular momentum of the group $G_i$, respectively, and the total angular momentum of the rest of the molecule is $J_0$. The total angular momentum of the molecule $\textbf{\textit{J}}$ is their vector sum, i.e.,
\begin{equation}
    \textbf{\textit{J}}=\textbf{\textit{J}}_0+\sum_{i=1}^2 \textbf{\textit{J}}_i.
\end{equation}
When VB interacts with the functional groups on the molecule, the angular momentum of one or both the groups could be changed. Therefore, the total angular momentum of the molecule would be
\begin{equation}
     \textbf{\textit{J}}'=\textbf{\textit{J}}_0+\sum_{i=1}^2 \textbf{\textit{J}}_i',
\end{equation}
with $J_i$ and $J_i'$ satisfy the selection rules Eq.(\ref{Tranrule}). As a result, the rotational transition of the big molecule could be analyzed by studying the rotational transition of its functional groups. 
\subsection{Interaction between molecules and non-canonical vortex beams}
\label{DiscB}
\begin{figure}
    \centering
    \includegraphics[scale=0.3]{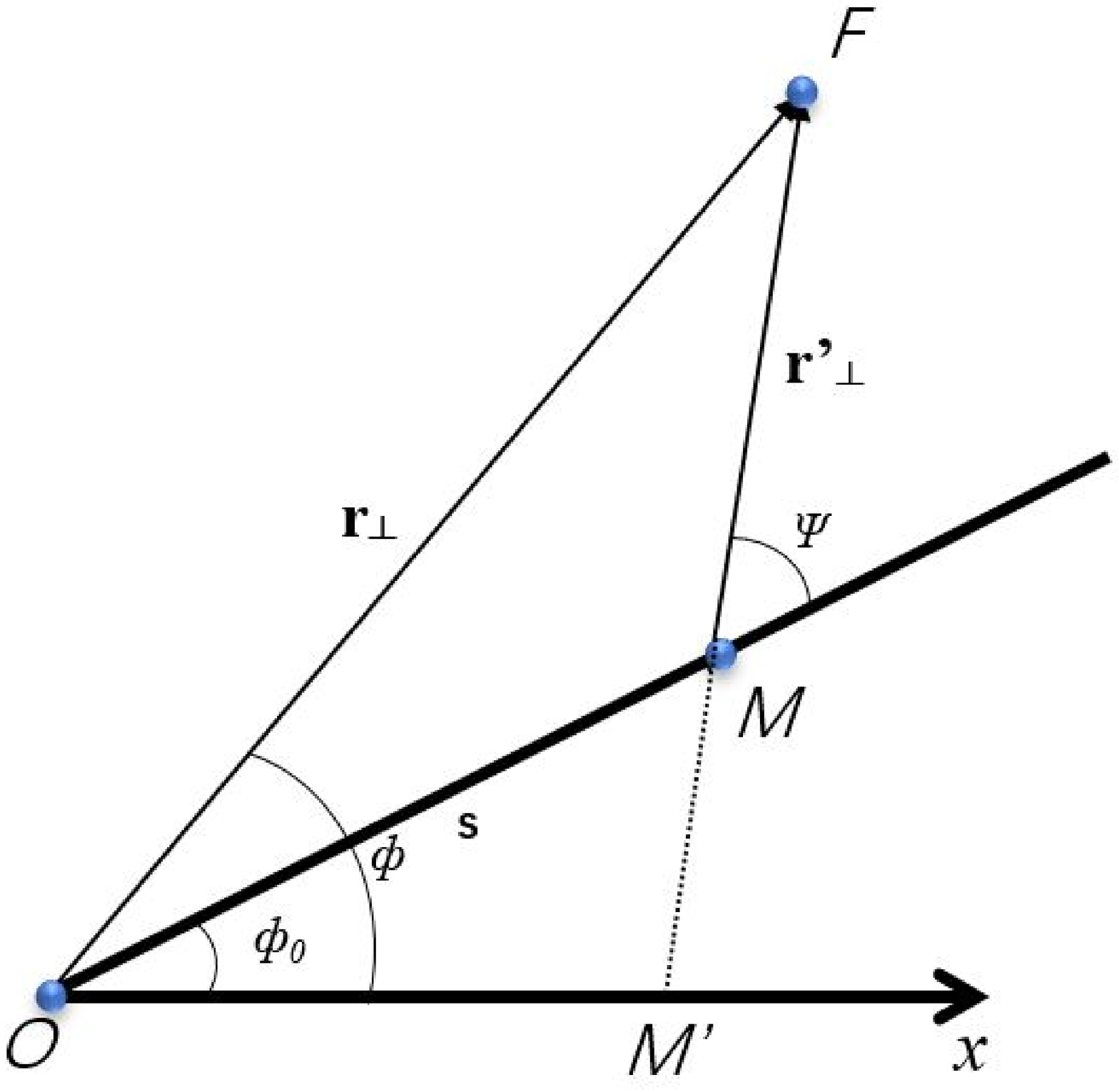}
    \caption{Extra geometry definition for the case of when molecule interacts with non-canonical vortex beam. On the transverse plane, the projections of the COM of the molecule and the active bound electron are points $M$ and $F$, respectively, and $O$ is the origin of space-fixed coordinates. The azimuthal angle in the phase factor is $\phi=\angle FOx$, and the azimuthal coordinate of the molecule's COM is $\phi_0=\angle MOx$. The definitions of the coordinates not showing up in this figure are the same as Fig.\ref{OMOA}.}
    \label{OVM2}
\end{figure}

One might also expect that, molecules could interact with non-canonical vortex beams \cite{molina2001propagation}, such as beams with Mathieu vortices \cite{loxpez2006orbital, chavez2002holographic}, power-exponential-phase vortex (PEPV) beams \cite{lao_propagation_2016, shen2019generation}, or sine-azimuthal wavefront modulated Gaussian (SWMG) beams \cite{lao_generation_2016}. The analysis given in this paper is generally also applicable to these problems with a few modifications in math. For these vortex beams, one way for obtaining the spatial distribution of electric field is to apply Fourier transform to the field expression and decompose it onto the orthogonal basis of the phase factor, i.e., $e^{il\phi}$, then calculate the propagation property for each component with the Collins formula so that the summation of these components could describe the total optical field. With this method, even though the non-canonical phase factors usually could not maintain its form of the mathematics expression in the process of propagation, the phase factor of each individual component is consistent during the propagation, and thus the electric field of the non-canonical vortex beams could be expressed in, or approximated to, the following form \cite{lao_propagation_2016}:  
\begin{equation}
\label{EMphi}
    E(r_\perp, \phi, z)=\displaystyle\sum_{l=-\infty}^{\infty} M_l(r_\perp, z) e^{il\phi},
\end{equation}
here the time-dependent factor is omitted. Since each component has an individual complex amplitude $M_l(r_\perp, z)$, the azimuthal symmetry of the intensity of the total field $I(r_\perp, \phi, z)=|E|^2$ might be missing, and thus the polar axis could not be defined as the ray that pointing to the COM of molecule. As a result, some extra geometry relationships should be taken into consideration, as shown in Fig.\ref{OVM2}.
The geometric relations give the approximation formula
\begin{eqnarray}
\label{aprx2}
    &&\phi \approx\phi_0+\frac{r'_\perp\mathrm{sin}\psi}{s},\\
    &&r_\perp \approx s+r'_\perp\mathrm{cos}\psi,\\
    &&r'_\perp=r'\mathrm{sin}\theta,\\
    &&z=z_0+r'\mathrm{cos}\theta,
\end{eqnarray}
On substitute these conditions into the Eq.(\ref{EMphi}), one can obtain the electric field expression in the vicinity of the point $(s, \phi_0, z_0)$ (here $\phi_0$ is the angular displacement from the origin of $\phi$ coordinate of the COM):
\begin{equation}
    E\approx F_0+r'\displaystyle\sum_{p=-1}^1C_pY_1^p(\theta, \psi),
\end{equation}
where 
\begin{equation}
    F_0=E(s,\phi_0, z_0)=\displaystyle\sum_{l=-\infty}^{\infty} M_l(s, z_0)e^{il\phi_0},
\end{equation}
\begin{equation}
\begin{split}
    C_{-1} =& \sqrt{\frac{2\pi}{3}}\displaystyle\sum_{l=-\infty}^{\infty}e^{il\phi_0}
    \bigg[\left.\frac{\partial M_l(r_\perp, z_0)}{\partial r_\perp}\right|_{r_\perp=s}\\
    &-\frac{iM_l(s_0, z_0)l}{s}\bigg],
\end{split}
\end{equation}
\begin{equation}
    C_0=\sqrt{\frac{4\pi}{3}}\displaystyle\sum_{l=-\infty}^{\infty}e^{il\phi_0}\left[\left.\frac{\partial M_l(s, z)}{\partial z}\right|_{z=z_0}\right],
\end{equation}
\begin{equation}
\begin{split}
    C_{1} =&-\sqrt{\frac{2\pi}{3}}\displaystyle\sum_{l=-\infty}^{\infty}e^{il\phi_0}
    \bigg[\left.\frac{\partial M_l(r_\perp, z_0)}{\partial r_\perp}\right|_{r_\perp=s}\\
    &+\frac{iM_l(s_0, z_0)l}{s}\bigg],
\end{split}
\end{equation}
and the rest of the calculation process of the transition amplitude is similar to the previous section and would be omitted here.

Further, although the summations have infinite terms, taking a truncation and summing over a finite range $[-l_{max}, l_{max}]$ might still let the finite sum approximate the electric field with high accuracy \cite{lao_propagation_2016}. This analysis is out of the scope of this paper and might be left as some future works.

\section{Conclusion}
\label{Conc}
In the preceding sections, we briefly reviewed the research of the interaction between vortex beams and molecules, and take a step forward by considering how the VB-molecule interaction could affect the diatomic molecular rotational states. To start, we first expanded the interaction Hamiltonian $H_{int}$ with spherical harmonics in the molecule-fixed coordinates with origins at the COM of the molecule, then calculated the transition amplitude for each pair of initial and final molecular states and discussed the selection rules for the transition. The result shows that, a photon of VB may carry one, two, or zero quanta of total angular momentum, and therefore it can drive different branches of molecular rotational transition during the interaction. For illustration, the simulation of the SrF molecular rotational spectrum of $B^2\Sigma^+\leftarrow X^2\Sigma^+$ system was given. We also briefly discussed the interaction between VB and larger molecules, as well as the case of when molecule interacts with non-canonical vortex beams. These theoretical results show that the VB-molecule interaction can be used for molecular state preparation as well as some further study. 

\begin{acknowledgments}
The author acknowledges the financial support from the Department of Physics and Astronomy, University of California, Los Angeles, thanks Prof. Eric R. Hudson for the helpful discussion about molecular states, and Prof. Daomu Zhao at Zhejiang University for the inspiring instruction about vortex beams. 
\end{acknowledgments}

\appendix
\section{Helicity basis and molecular rotational states}
\label{helib}
\subsection{Light-atom interaction and Helicity basis}
In this paper, we use helicity basis to express the Hamiltonian. For illustration about the helicity basis, we can start with a well-known example: light-atom interaction. When a laser with electric field strength $E$ and polarization $\hat{\epsilon}$ interacts with an atom, the photon could drive the dipole transition between two electronic states, with its SAM participate in the angular momentum transfer process. For example, as a ground state alkaline atom (Na, K, Rb, etc.) is illuminated by a laser which could drive its outer electron at ground state $|n^2S_{1/2}\rangle$ to the excited state $|n^2P_{3/2}\rangle$, the transition amplitude between angular momentum substates $|n^2P_{3/2}, m_l'\rangle\leftarrow|n^2S_{1/2}, m_l=0\rangle$ is
\begin{equation}
    A=\langle n^2P_{3/2}, m_l'|E(-e)\hat{\epsilon}\cdot\textbf{\textit{r}}|n^2S_{1/2}, m_l\rangle,
\end{equation}
where $n, l$ and $m_l$ are the principle, angular and magnetic quantum numbers, respectively, $\textbf{\textit{r}}$ is the position vector of the electron with respect to the nucleus. According to the spherical symmetry, the electronic wave functions could be expressed as the products of radial and angular parts, i.e., $|\varphi_{n, l, m_l}(\textbf{\textit{r}})\rangle=R_{n,l}(r)Y_l^{m_l}(\theta, \phi)$. Then, one can express the interaction Hamiltonian in the helicity basis and carry out the calculation. The relations between the helicity and Cartesian unit vectors are
\begin{equation}
    \textbf{\textit{e}}_{\pm1}=(-\textbf{\textit{e}}_x\pm i\textbf{\textit{e}}_y)/\sqrt{2}, \textbf{\textit{e}}_0=\textbf{\textit{e}}_z.
\end{equation} 
For simplicity, we would assume the laser to be circularly polarized, i.e., $\hat{\epsilon}=\textbf{\textit{e}}_{+1}$ or $\textbf{\textit{e}}_{-1}$ (and the linear polarization could be expressed as a linear combination of $\textbf{\textit{e}}_{+1}$ and $\textbf{\textit{e}}_{-1}$). With the definition, the dot product $\hat{\epsilon}\cdot\textbf{\textit{r}}$ could therefore be written as
\begin{equation}
    \hat{\epsilon}\cdot\textbf{\textit{r}}=r\sqrt{\frac{4\pi}{3}}Y_1^{\pm1}(\theta, \phi),
\end{equation}
where the helicity basis expression for the position operator
\begin{equation}
    \textbf{\textit{r}}=x\textbf{\textit{e}}_x+y\textbf{\textit{e}}_y+z\textbf{\textit{e}}_z=r\sum_{p=-1}^1\sqrt{\frac{4\pi}{3}}Y_1^p(\theta, \phi)\textbf{\textit{e}}_p^*
\end{equation}
has been used. For a given polarization $p$, the transition amplitude reads
\begin{equation}
\begin{split}
    A&=E(-e)\sqrt{\frac{4\pi}{3}}\int_0^\infty R_{3, 1}(r)R_{3, 0}(r)r^3\mathrm{d}r \\
    &\times \int_0^\pi\mathrm{sin}\theta\mathrm{d}\theta\int_0^{2\pi}\mathrm{d}\phi Y_1^{m_l}(\theta, \phi)^*Y_1^p(\theta, \phi)Y_0^0(\theta, \phi).
\end{split}
\end{equation}
The radial integral in the first line is not of our main focus. For the angular part, we apply the spherical harmonics integral formula \cite{brown_rotational_nodate} and obtain
\begin{equation}
\begin{split}
    &\int_0^\pi\mathrm{sin}\theta\mathrm{d}\theta\int_0^{2\pi}\mathrm{d}\phi Y_1^{m_l}(\theta, \phi)^*Y_1^p(\theta, \phi)Y_0^0(\theta, \phi)\\
    &=(-1)^{1-m_l'}\sqrt{3}\begin{pmatrix}1&1&0\\-m_l'&p&0\end{pmatrix}.
\end{split}
\end{equation}
Here the non-zero conditions for the Wigner 3-j symbol are actually the selection rules for the transition. The result shows that using helicity basis and expressing the Hamiltonian in spherical harmonics could bring some convenience in describing the selection rules.
\subsection{Molecular rotational states}
As we study the interaction between light and a bound electron in a molecule, the analysis is complicated because molecule usually does not have spherical symmetry and its electronic wave functions could not be expressed as the product of radical and angular parts as the atomic cases. For symmetric top molecule, its rotation could be described by the symmetric top wave function \cite{brown_rotational_nodate}
\begin{equation}
\label{JOMJ}
\begin{split}
    |J\Omega M_J\rangle=&\left[\frac{2J+1}{8\pi^2}\right]^{\frac{1}{2}}D_{\Omega M_J}^J(\alpha, \beta, \gamma)^*\\
    =&\left[\frac{2J+1}{8\pi^2}\right]^{\frac{1}{2}}e^{i\Omega\alpha}d_{\Omega M_J}^J(\beta)e^{iM_J\gamma},
\end{split}
\end{equation}
where $D_{\Omega M_J}^{J}(\alpha, \beta, \gamma)$ is the $\Omega M_J$ element of the rotation matrix $\mathcal{D}^{(J)}(\omega)$ with Euler angles $\omega=(\alpha, \beta, \gamma)$. It is easy to verify that this is the eigenstate of operators $-i\partial/\partial\alpha$ and $-i\partial/\partial\gamma$, and thus we can define them respectively as $J_z$ and $J_\zeta$. 

Further, for the linear molecule (such as the diatomic molecule in our case), since the rotational degree of freedom about the molecular axis is missing, the rotational wave function reads
\begin{equation}
    |J\Omega M_J\rangle=\left[\frac{2J+1}{4\pi}\right]^{\frac{1}{2}}D_{\Omega M_J}^J(0, \beta, \gamma)^*,
\end{equation}
in which the phase factor about axis $\zeta$, $e^{i\Omega\alpha}$, is missing. Therefore, the non-zero transition amplitude Eq.(\ref{qtca}) is only contributed by the $q=0$ components, which means only transitions with $\Delta \Omega=0$ are allowed.
\section{Coefficients of Spherical Harmonics in the expression of $H_{int}$}
\label{ApB}
Since the interaction Hamiltonian has two parts, $H_d$ and $H_{quad}$, the spherical harmonics coefficients could be calculated by summing their individual contribution. For the quadrupole interaction Hamiltonian $H_{quad}$, according to the Eq.(\ref{Hquad}), the spherical harmonics expansion is
\begin{equation}
\label{Hqsph}
\begin{split}
    H_{quad}=&\displaystyle\sum_{p=-2}^2\frac{(-1)^p}{2}T_{-p}^2(\nabla \mathbf{E})T_{p}^2(Q)\\
    =&\sum_{p=-2}^2\bigg[\frac{(-e)(-1)^p}{2}\sqrt{\frac{16\pi}{5}}T_{-p}^2(\nabla \mathbf{E})\bigg]r'^2Y_2^p(\theta, \psi),
\end{split}
\end{equation}
and it is obvious to see that the terms in the square brackets contribute to the expansion coefficients, $a_{k, p}^{(\pm)}$. Note that we have neglected the interaction terms with orders of $r'$ higher than Eq.(\ref{Eapprox}), and in $H_{quad}$, the highest order of $r'$ in Eq.(\ref{Hqsph}) is decided by $T_p^2(Q)$, therefore the rank-2 tensor expression for the components of the gradient of the helical electric field, $T_p^2(\nabla \mathbf{E^{(\pm)}})$, could be approximated to the first order of $r'$:
\begin{eqnarray}
    &&T_0^2(\nabla \mathbf{E^{(\pm)}})\approx\frac{\sqrt{3}}{6}\left[\pm\frac{lF_0}{s}-F_\alpha\right],\\
    &&T_1^2(\nabla \mathbf{E^{(\pm)}})\approx \frac{\sqrt{2}}{4}\left[-F_\beta\pm F_\beta\right],\\
    &&T_2^2(\nabla \mathbf{E^{(\pm)}})\approx -\frac{\sqrt{2}(-1\pm1)}{4}\left[F_\alpha - \frac{lF_0}{s}\right],\\
    &&T_{-1}^2(\nabla \mathbf{E^{(\pm)}})\approx\frac{\sqrt{2}}{4}\left[F_\beta\pm F_\beta\right],\\
    &&T_{-2}^2(\nabla \mathbf{E^{(\pm)}})\approx\frac{\sqrt{2}(1\pm1)}{4}\left[F_\alpha + \frac{lF_0}{s}\right],
\end{eqnarray}
where $(\pm)$ in the superscript is to distinguish the polarization of the beam, and $\mathbf{E^{(\pm)}}=E\mathbf{\hat{\epsilon}^{\pm}}$. 

To evaluate the contribution of $H_d$, we first express it in form of spherical harmonics:
\begin{equation}
\begin{split}
    H_d=&E\mathbf{\hat{\epsilon}^{\pm}\cdot d}\\
    =&\left[F_0+r'\displaystyle\sum_{p=-1}^1 C_p Y_1^p(\theta,\psi)\right]\\
    &\times (-e)r'\sqrt{\frac{4\pi}{3}}Y_1^{\pm1}(\theta, \psi)\\
    =&\sum_{k=0}^2\sum_{p=-k}^kA_{k, p}^{(\pm)}r'^{n_k}Y_k^p(\theta, \psi),
\end{split}
\end{equation}
where $n_k=1$ when $k=1$ and $n_k=2$ for $k=0$ and $k=2$, and
\begin{eqnarray}
&&A_{0,0}^{(\pm)}=\frac{\sqrt{2\pi}}{3}\left[\frac{lF_0}{s}\mp F_\alpha\right],\\
&&A_{1, 0}^{(\pm)}=A_{1, 1}^{(-)}=A_{1, -1}^{(+)}=0,\\
&&A_{1, 1}^{(+)}=A_{1, -1}^{(-)}=\sqrt{\frac{4\pi}{3}}F_0,\\
&&A_{2, 0}^{(\pm)}=\frac{\sqrt{10\pi}}{15}\left[\pm F_\alpha-\frac{lF_0}{s}\right],\\
&&A_{2,1}^{(+)}=A_{2,-1}^{(-)}=\sqrt{\frac{4\pi}{15}}F_\beta,\\
&&A_{2,2}^{(+)}=\sqrt{\frac{4\pi}{15}}\left[- F_\alpha-\frac{lF_0}{s}\right],\\
&&A_{2,-2}^{(-)}=\sqrt{\frac{4\pi}{15}}\left[F_\alpha-\frac{lF_0}{s}\right],\\
&&A_{2,-1}^{(+)}=A_{2,1}^{(-)}=A_{2,-2}^{(+)}=A_{2,2}^{(-)}=0.
\end{eqnarray}
Here the formula for the product of spherical harmonics functions has been used. By adding the coefficients together, we have
\begin{eqnarray}
    &&a_{2,p}^{(\pm)}=A_{2,p}^{(\pm)}+\frac{(-1)^p}{2}T_{-p}^2(\nabla \mathbf{E^{(\pm)}}),\\
    &&a_{1,p}^{(\pm)}=A_{1,p}^{(\pm)},\\
    &&a_{0,0}^{(\pm)}=A_{0,0}^{(\pm)}.
\end{eqnarray}

\newpage

\bibliography{Reference}

\end{document}